\documentclass[useAMS,usenatbib]{mn2e}
\usepackage{graphics}
\usepackage{amsmath}
\usepackage{amssymb}

\def\apj{ApJ}

\def\apjl{ApJL}
\def\mnras{MNRAS}

\def\nat{Nature}

\def\apjs{ApJS}

\def\prd{Physical Review D}

\newcommand{\ea}{et al.}
\newcommand{\lta}{\lesssim}

\newcommand{\kpc}{\>{\rm kpc}}
\newcommand{\Mpc}{\>{\rm Mpc}}

\newcommand{\hinv}{\>{h^{-1}}}
\newcommand{\msol}{\>{\rm M_{\sun}}}

\newcommand{\bdm}{\begin{displaymath}} 
\newcommand{\edm}{\end{displaymath}}
\newcommand{\beq}{\begin{equation}} 
\newcommand{\eeq}{\end{equation}} 
\newcommand{\bit}{\begin{itemize}} 
\newcommand{\eit}{\end{itemize}} 
\newcommand{\ben}{\begin{enumerate}} 
\newcommand{\een}{\end{enumerate}}
\newcommand{\bfi}{\begin{figure}[htb]} 
\newcommand{\bpfi}{\begin{figure}[p]}

\newcommand{\GO}{\texttt{GO} }
\newcommand{\PH}{\texttt{PH} }
\newcommand{\CSF}{\texttt{CSF} }
\newcommand{\ficm}{$f_{\rm ICM}$} 
 
\newcommand{\rfiveh}{$r_{500}$}

\title[The Effect of Gas Physics on the Halo Mass Function]{The Effect of Gas Physics on the Halo Mass Function}
\author[R. Stanek, D. Rudd and A. E. Evrard]{R. Stanek$^{1}$\thanks{E-mail:
rstanek@umich.edu}, D. Rudd$^{2}$, and A. E. Evrard$^{1,3}$\\
$^{1}$Department of Astronomy, University of Michigan, 500 Church St., Ann Arbor, MI  48109\\
$^{2}$School of Natural Sciences, Institute for Advanced Study, Einstein Drive, Princeton, NJ  08540\\
$^{3}$Department of Physics and Michigan Center for Theoretical Physics, University of Michigan, 450 Church St., Ann Arbor, MI  48109}
\begin{document}

\date{}

\maketitle


\begin{abstract}
Cosmological tests based on cluster counts require accurate calibration of the space density of massive halos, but most calibrations to date have ignored complex gas physics associated with halo baryons.   We explore the sensitivity of the halo mass function to baryon physics using two pairs of gas-dynamic simulations that are likely to bracket the true behavior.   Each pair consists of a baseline model involving only gravity and shock heating, and a refined physics model aimed at reproducing the observed scaling of the hot, intracluster gas phase.   One pair consists of billion-particle re-simulations of the original $500 \hinv\Mpc$ Millennium Simulation of Springel \ea (2005), run with the SPH code Gadget-2 and using a refined physics treatment approximated by preheating (PH) at high redshift.   The other pair are high-resolution simulations from the adaptive-mesh refinement code ART, for which the refined treatment includes cooling, star formation, and supernova feedback (CSF).   We find that, although the mass functions of the gravity-only (GO) treatments are consistent with the recent calibration of  \cite{tinker:08}, both pairs of simulations with refined baryon physics show significant deviations.  Relative to the GO case, the masses of $\sim 10^{14} \hinv\msol$ halos in the PH and CSF treatments are shifted by averages of $-15 \pm 1$ percent and $+12 \pm 5$ percent, respectively.   These mass shifts cause $\sim \! 30\%$ deviations in number density relative to the Tinker function, significantly larger than the $5\%$ statistical uncertainty of that calibration.   
\end{abstract}


\small

\section{Introduction}

Deep cluster surveys offer the promise of tightly constraining cosmological parameters, including the nature of dark energy  \citep{holder:01,levine:02,majumdar:03,lima:04,lima:05,younger:06,sahlen:08}.  Realizing this promise requires accurate calibration of the expected counts and clustering of massive halos, along with a  careful treatment of how halo mass relates to the signals observed by such surveys.  This logical division is reflected by two long-standing threads of effort, one focused on the emergence of massive structures from gravity and the other focused on scaling relations of multiple signals within the population of massive halos.   

The fact that $17\%$ of clustered matter in the universe is baryonic ties these threads together.  Non-gravitational physics is required in massive halos, not simply to create galaxies \citep{white:78} but also to reproduce scaling behavior of the hot, intracluster medium (ICM)  observed in X-rays \citep{evrard:91,borgani:01,reiprich:02,stanek:06,nagai:07}.  If a significant fraction of halo baryons become spatially segregated from the dark matter, either condensed within galaxies or disbursed from non-gravitational heating, then the gravitational development of massive structures will be altered, perhaps at the $\sim 10\%$ level, under strong baryonic effects.  

The spatial number density of halos, or mass function, expected from Gaussian random initial conditions was originally derived using a mix of analytic arguments and numerical simulations \citep[e.g.][]{ps:74,bond:91,st:99}.  Modern efforts focus on providing fitting functions of increasing statistical precision  \citep{jenkins:01,warren:06,tinker:08}.  The recent \citet{tinker:08} mass function (hereafter TMF), calibrated to a wide range of cosmological simulations that include gas-dynamic, Marenostrum simulations \citep{yepes:07,gottlober:07}, has pushed statistical errors to the level of $5\%$.  

To date, however, there have been few gas-dynamic simulations 
that include a non-gravitational treatment of baryonic processes in volumes large enough to provide good statistics for high-mass halos.  Calibration of the mass function at the level of
the \citet{tinker:08} using hydrodynamic simulations is too expensive to be feasible 
in the near-term.  A less computationally expensive technique is to compare realizations of fixed initial conditions evolved with different baryonic physics.  \citet{jing:06} and \citet{rudd:08} employ this approach to study baryonic effects on the matter power spectrum,  finding 2-10\% modifications of the matter power spectrum at scales $k \sim 1 h \Mpc^{-1}$.  \cite{rudd:08} finds a halo mass function that 
is enhanced by $\sim10\%$ relative to the dark-matter only case.  Neither set of simulations
were sufficiently large to properly probe rich cluster scales, however.

In this letter, we take a similar approach to examine the effect of non-gravitational, baryonic physics 
on the cluster mass function.  Specifically, we consider two pairs of gas-dynamic simulations, each comprised of a treatment of the gas with gravity and shock heating only and a second, more complicated treatment.   One pair are Millennium Gas Simulations \citep[MGS hereafter]{hartley:08}, with SPH gas dynamics under Gadget-2 \citep{gadget2}, and the second pair are adaptive-mesh ART simulations from \cite{rudd2007phd}.    As the two simulations in each pair have the same initial conditions, we infer the effects of baryonic physics by comparing halos in the more detailed simulations with their gravity-only counterparts.
 The outline of this paper is as follows: in Section \ref{sec:sims}, we discuss the various simulations and their bulk cluster properties, comparing them to observations.  Section \ref{sec:mf} compares halo masses between corresponding halos in each pair of simulations, and discusses the effect 
on the total mass function.  All halos masses are identified as $M_{500}$, the mass of a spherical halo 
with radius $r_{500}$ and mean density $500\rho_c(z)$, where $\rho_c(z)$ is the critical density of the 
universe.

\section{Simulations and Halo Samples} \label{sec:sims}

We use two pairs of gas dynamic simulations, each pair 
run from a single set of initial conditions and differing only in the included physical
processes.  Our baseline simulations are evolved with only gravity and shock heating acting on the gas (hereafter \texttt{GO}) and a second has more complicated treatment of the gas physics discussed below.  

Halos in all simulations are identified as spherical regions, centered on the peak of the 
dark matter distribution, where the mean enclosed density is $500 \rho_c(z)$.  Our analysis focuses on $z =0$ and $z=1$ samples, redshifts that roughly bracket the range important for dark energy studies.   When calculating the bulk cluster properties, we exclude from our sample halos that overlap with more massive neighbors; however, to remain consistent with \cite{tinker:08} halos that overlap but whose centers lie outside their respective virial radii are included in the mass function analysis.

\subsection{Millennium Gas Simulation} \label{sec:msg}

The Millennium Gas Simulations (hereafter MGS) are a pair of resimulations of the 
 Millennium Simulation \citep{springel:05}, a high-resolution, dark-matter-only simulation 
of a $500 \hinv\Mpc$  cosmological volume. Like the original Millennium Simulation, the simulations were run with GADGET-2,  which treats the gas dynamics with smoothed particle hydrodynamics (SPH) 
\citep{gadget2}.  The MGS runs use a down-sampled version of the initial  
conditions of the Millennium simulation, with $5 \times 10^8$ dark matter particles, each of 
mass $1.422 \times 10^{10} \hinv\msol$, and $5 \times 10^8$ SPH gas particles, each of 
mass $3.12 \times 10^9 \hinv\msol$.  This mass resolution is about 20 times coarser 
than the original Millennium simulation, and the gravitational softening length of $25 \hinv\kpc$ is correspondingly larger.  The 
cosmological parameters match the Millennium Simulation: ($\Omega_m$, $\Omega_b$, $\Omega_\Lambda$, $h$, $n$, $\sigma_8$) =
(0.25, 0.045, 0.75, 0.73, 1.0, 0.9).  

Complementing the aforementioned \GO realization is an MGS simulation 
with cooling and preheating, denoted as \PH.  Preheating is a simple approximation that assumes high redshift galaxy formation feedback drove the proto-ICM gas to a fixed entropy level, after which the ICM evolves under hierarchical gravity \citep{evrard:91,kaiser:91,bialek:01}.  In our implementation, the entropy of each gas particle is instantaneously boosted to 200 keV cm$^2$ at $z = 4$.   The gas is allowed to radiatively cool thereafter using the cooling function of \cite{sutherland:93}, but the cold gas fraction is very small.  The entropy level of the \PH model is tuned to match bulk X-ray observations of clusters at redshift zero, as we discuss shortly.

For both MGS models, we calculate bulk cluster properties with primary halos of mass $M_{500} \ge 5 \times 10^{13} \hinv\msol$, yielding sample sizes of 2527 (\texttt{PH}) and 3446 (\texttt{GO}) at $z = 0$ and of 475 (\texttt{PH}) and 818 (\texttt{GO}) at $z = 1$.   

\subsection{ART Simulations}

Our second set of models are simulated using the 
distributed-parallel hydrodynamic ART code \citep{rudd:08,rudd2007phd}.   The simulations 
evolve a $240^3 h^{-3} \mathrm{Mpc}^3$ volume of a WMAP3-motivated cosmological model with 
parameters, ($\Omega_m$, $\Omega_b$, $\Omega_\Lambda$, $h$, $n$, $\sigma_8$) =
(0.25, 0.042, 0.75, 0.73, 0.95, 0.8).  The baseline \GO simulation was performed
using $512^3$ dark matter particles with mass $m_p \sim 5.95\times10^{9} \hinv\msol$ and allow 
for 4 levels of refinement achieving a minimum cell size of 
$240 \hinv\Mpc/(512\times2^4) \approx 29 \hinv\kpc$.  We then
selected the 13 most massive halos in the simulation volume at $z = 0$, and
resimulated at $1024^3$ effective resolution ($m_p \approx 7.44\times10^{8} \hinv\msol$) the 
regions within $5 r_{\mathrm{vir}} \sim 5-10 \hinv\Mpc$ surrounding each cluster center.  This 
simulation includes prescriptions for star formation and metal-dependent radiative cooling 
described in \citet{rudd2007phd}.   For this simulation, the same $512^3$ uniform mesh was used, 
but in the high-resolution regions 7 levels of refinement were used for a peak spatial resolution 
of $\approx3.6 \hinv\kpc$.


\subsection{Baryon Census}\label{sec:baryons}

We begin by exploring the bulk properties of baryons in the massive halo samples as a means of assessing the viability of the different physical treatments. 
  
\begin{figure}
\scalebox{0.38}{\includegraphics{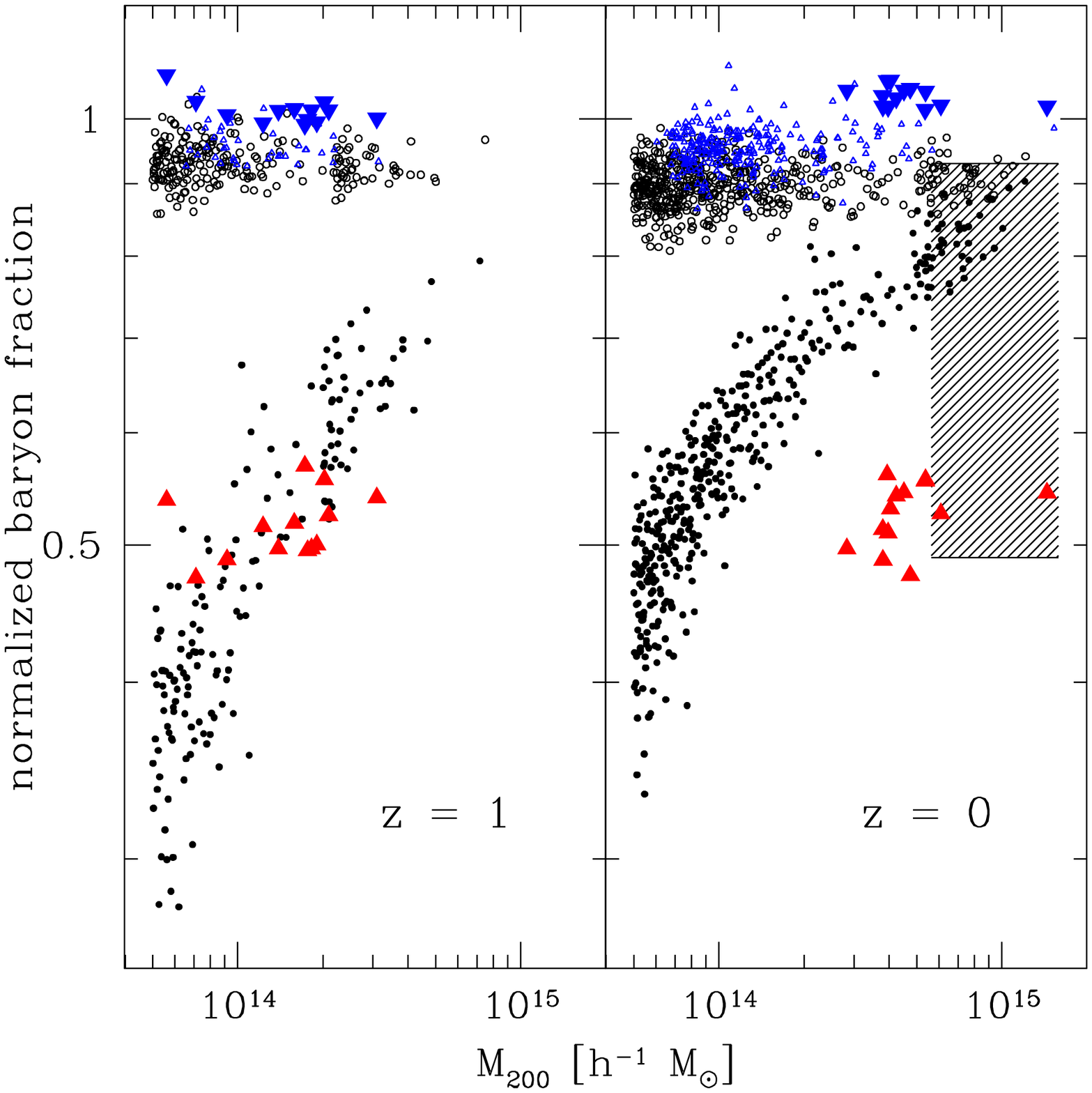}}
\caption{Baryon fractions as a function of total halo mass at $z=1$ (left panel) and $z=0$ (right panel) are shown for the MGS halo samples in the \GO (open circles) and \PH (filled circles) treatments and for the ART samples in the \GO (open triangles) and \CSF (filled triangles) cases.   For the latter, regular triangles show the ICM mass fraction while inverted triangles show the total baryon fraction (gas plus stars) within the halos.   The shaded region shows the $90\%$ confidence range of the mean, observed ICM mass fraction inferred for local, $kT > 4\,$keV clusters by \citet{vikhlinin:06}.   
For clarity, only a subset of the MGS samples are shown for masses below $4~\times~10^{14} \hinv\msol$.}
 \label{fig:ficm}
\end{figure}

Figure~\ref{fig:ficm} shows baryon mass fractions within \rfiveh, normalized to the universal baryon fraction $\Omega_b/\Omega_m$.   In the \GO simulations, all the baryons are in the hot intracluster medium phase, so that $f_b = f_{\rm ICM}$.  In the \CSF simulation, gas is removed from the hot phase through radiative cooling and converted to stars.  For these halos, we plot both the ICM mass fraction \ficm\ and the total baryon fraction, $f_b = f_{\rm ICM} + f_{\rm cond}$, where condensed baryons, $f_{\rm cond}$, includes both stars and cold gas ($T < 2\times10^5 K$).  Although the \PH model allows radiative cooling, the fraction of cold gas in our halo samples is very small, less than two percent of the baryons.

The \GO simulations display constant baryon fractions that are slightly suppressed from the universal mean value.  The level of suppression is somewhat larger in the MSG halos compared to the ART sample, which is  consistent with the difference between SPH and grid codes reported by \citet{kravtsov:05}. The mass-limited samples have average baryon fractions at $z=0$ of $f_b = 0.89 \pm 0.025$ and  $0.93 \pm 0.038$, respectively.  The MGS baryon fractions are consistent with those in SPH simulations done at similar resolution and measured within $\Delta_c = 200$ by \cite{crain:07} and \cite{ettori:06}.

The baryon distributions in the \PH and \CSF simulations are more complicated.  In the \PH case, the entropy increase at $z \sim 4$ causes the gas to expand, especially in lower-mass halos for which the characteristic entropy is lower, and raises the sound speed throughout the proto-ICM.  The latter effect pushes the effective shock radius to larger values compared to the purely gravity-driven case \citep{voit:03}.  As a result, lower-mass halos retain a smaller fraction of their baryons within \rfiveh, leading to the mass-dependence seen in Figure \ref{fig:ficm}.  Since the characteristic halo entropy increases with time at fixed mass, the mean ICM gas fraction at fixed mass increases from $z=1$ to $z=0$ in the \PH halos.   At $z = 0$, the highest mass halos have baryon fractions suppressed by only 10\% relative to the \GO treatment.

In the \CSF halos, the hot gas fraction, \ficm, is comparable to that in the \PH clusters of similar mass at $z = 1$.   However, the total baryon fraction in these halos is close to universal, due to the contribution of cold gas and stars.  Unlike the \PH models, the ICM mass fraction does not evolve with time, remaining approximately constant at $\approx50\%$ from $z = 1$ to $z = 0$, even as the clusters themselves grow by a factor of two in total mass.  The total baryon fraction grows by 4\% due primarily to the small increase in the ICM.  The stellar component grows significantly from $\sim40\%$ to $\sim47\%$ but is balanced by a corresponding decrease in the fraction of cold gas from $\sim10\%$ to $\sim4\%$.

\begin{figure}
\scalebox{0.38}{\includegraphics{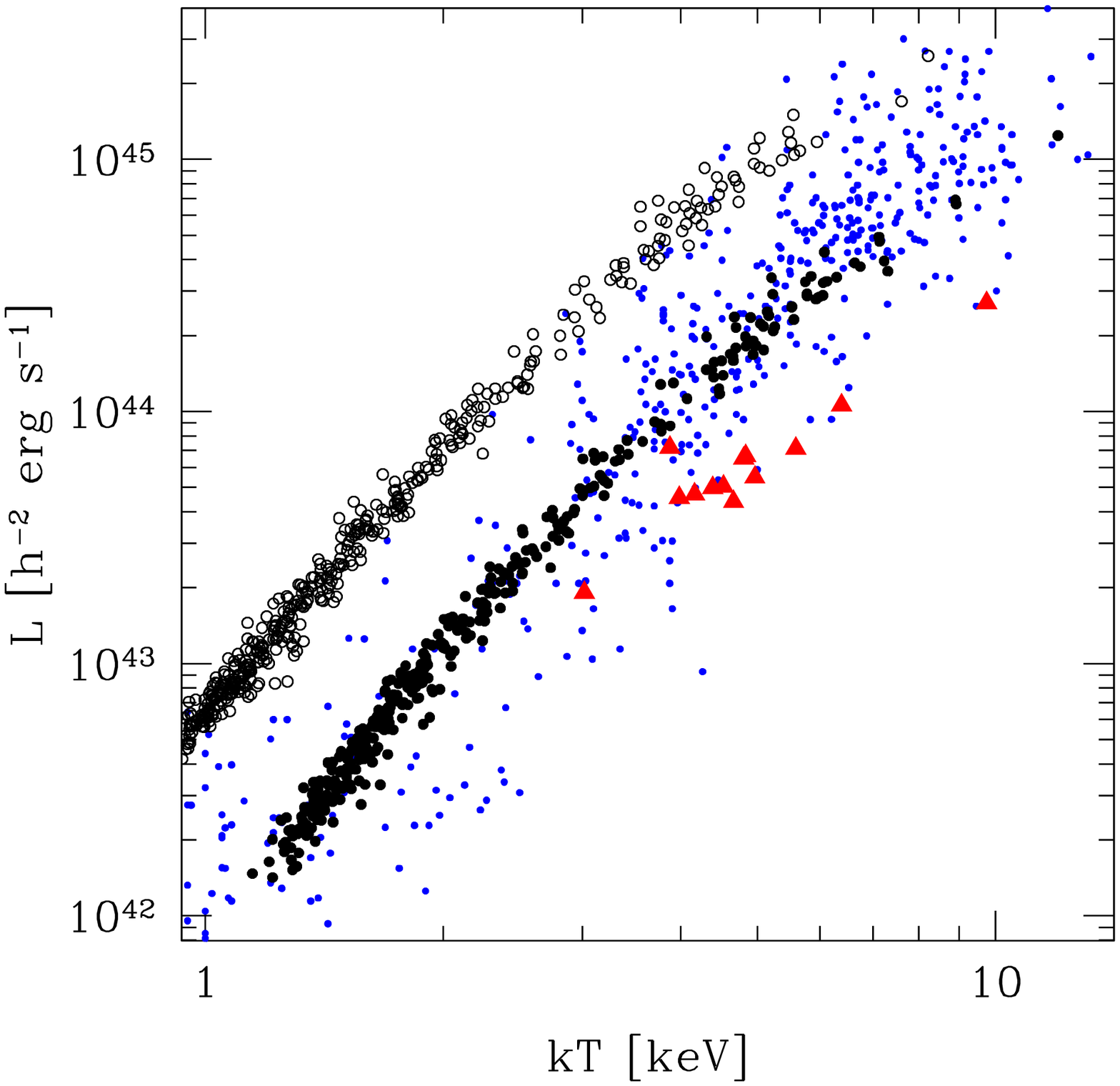}}
\caption{The $L_{\rm bol}-T_{\rm sl}$ relations for the MGS--\GO (open cirles), MGS--\PH (filled circles), and ART--\CSF (filled triangles) samples are compared to observations (small points) compiled by \citet{hartley:08}.   The MGS halos are sub-sampled as in Figure~\ref{fig:ficm}.  }
\label{fig:elt}
\end{figure}

As an empirical test of the models, we show in Figure \ref{fig:elt} the scaling between bolometric luminosity $L_{\rm bol}$  and spectral temperature $T_{\rm sl}$ for the $z=0$ halo samples of the MGS and ART--\CSF simulations.  The models are compared to a local sample of clusters compiled by \cite{hartley:08}.   As the local sample extends only to modest redshifts, $z \lta 0.2$, we do not apply  evolutionary corrections to the observations.

For the models, we use the analytic approximation of \cite{bartelmann:96} to compute $L_{bol}$ within $r_{200}$ of each halo.  For the \GO and
\PH halos, we compute spectroscopic temperatures, $T_{\rm sl}$, using the expression in \cite{mazzotta:04}.  This expression is known to fail at low temperatures, so, for the \CSF clusters, we use instead the method of \citet{vikhlinin:06}.  Additionally, for the \CSF clusters we exclude gas within $0.1 r_{200}$ and within dark matter substructures to crudely reproduce the clump removal procedure applied in \citet{nagai:07} and \citet{rasia:06}.  Applying this simple analysis proceedure to the simulated clusters used in \citet{nagai:07} give temperatures that differ by $\sim10\%$ or less from the mock {\it Chandra} analysis.  The measured X-ray quantities are sensitive to the choice of innermost radius, with larger cuts leading to simultaneously lower measured $L_{\rm bol}$ and $T_{\rm sl}$.  

Both of the non-gravitational physics models provide a better match to the observed data than the \GO simulation.  As discussed in \cite{hartley:08}, the \PH halos match the slope and normalization of the observed $L-T$ relation well.   The observed scatter is much larger, however, due primarily to the existence of cool cores in real clusters.  

The slope of the \CSF halos also agrees with the observations, but the normalization and scatter are not well matched to the data.  The normalization offset is partly due to the lower gas fraction seen in Figure~\ref{fig:ficm}, but the spectral temperatures also play a role.  As discussed in \cite{borgani:04} and \cite{nagai:07}, the temperature profile of the hot phase is steeper than observed in cluster cores, resulting in enhanced $T_{\rm sl}$ values.

In summary, we have shown that both the \PH and \CSF simulations offer a reasonable match to the form of the $L-T$ relation, but the overall baryon content of halos differs substantially between the two.  In the \CSF simulation, star formation is overly efficient, so that nearly 50$\%$ of the baryons are in stars rather than in the  hot phase.  In the  \PH simulation, the stellar fraction is entirely neglected, but the net heating effect of early galaxy formation is assumed to be large enough to drive the halo baryon fraction substantially below the global value.   Neither of these treatments is fully consistent with observations, but they represent two extreme approximations for the true behavior.  We next examine the effects that these treatments have on halo mass.

\section{Halo Masses and the Mass Function}\label{sec:mf}

\begin{figure}
\scalebox{0.38}{\includegraphics{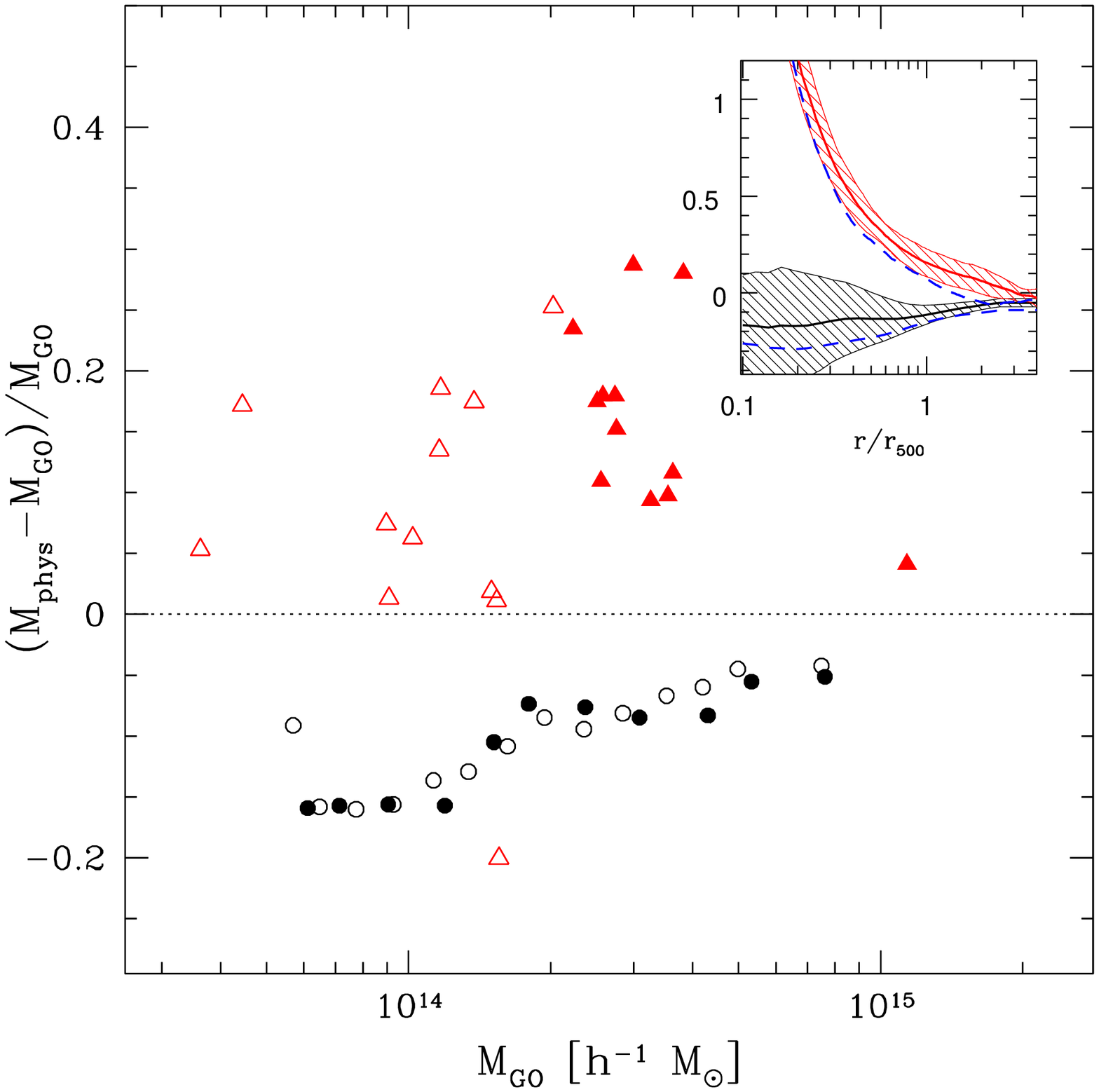}}
\caption{Fractional mass difference in halo mass with respect to the \GO realization.  Circles show the mean shift for the \PH halos at $z = 0$ (filled) and $z=1$ (open).  Triangles show individual  \CSF halos at $z = 0$ (filled) and $z=1$ (open).  The inset panel plots the cumulative radial mass difference for \CSF halos (red) and \PH halos (black) at $z=0$ (solid, with 1$-\sigma$ scatter) and $z=1$ (dashed).}
\label{fig:dm}
\end{figure}

Since both pairs of simulations are evolved from the same initial conditions, we are able to match halos between the realizations performed under the two physical treatments.   In Figure~\ref{fig:dm}, we show the fractional shift in mass that occurs under the \PH and \CSF treatments, relative to the respective \GO model, as a function of \GO halo mass at redshifts $z=0$ and 1.    The mean mass shift are plotted for MGS halos in mass bins.  Individual clusters are plotted for the ART simulations at $z = 0$ and $z = 1$.  

The \PH halos experience a substantial decrease in mass relative to the \GO treatments.  The magnitude of the $z=0$ fractional mass shift depends on halo mass, declining from $15\%$ at $10^{14} \hinv\msol$ to $5\%$ at  $10^{15} \hinv\msol$.   Although these mass shifts are mostly due to the change in gas fraction,  there is also a difference in dark matter structure that enhances the shift.  
All but the most massive ART--\CSF halo show increased mass relative to the respective \GO halos.  At the mean mass of the sample at $z = 0$, $3 \times 10^{14} \hinv \msol$, the mean fractional mass shift is $0.117 \pm 0.015$, including the outlier data point.  Approximately $2\%$ of this shift is due to the increase in baryon mass.  The remainder is due to the change in halo structure brought about by baryon cooling \citep{gnedin:04,nagai:06}.  

These mass shifts depend on the choice of scale, as shown in the inset of Figure \ref{fig:dm}.  For comparison with the ART--\CSF halos we plot the mean mass profile for MGS halos in the range $1-3 \times 10^{14} \hinv\msol$.  Within the core, the mass difference between matched halos in the MGS simulations is nearly $20\%$, but the mass difference approaches zero on scales significantly larger than $r_{200}$.  In the ART simulations, we also see that the mass shift is a strong function of scale: within the core it is very high, $\sim 80\%$, and approaches zero beyond $r_{200}$.  Because of this scale dependence, the magnitude of the mean mass shift and its evolution with redshift is sensitive to our choice of $\Delta = 500 \rho_c(z)$.

\begin{figure}
\scalebox{0.38}{\includegraphics{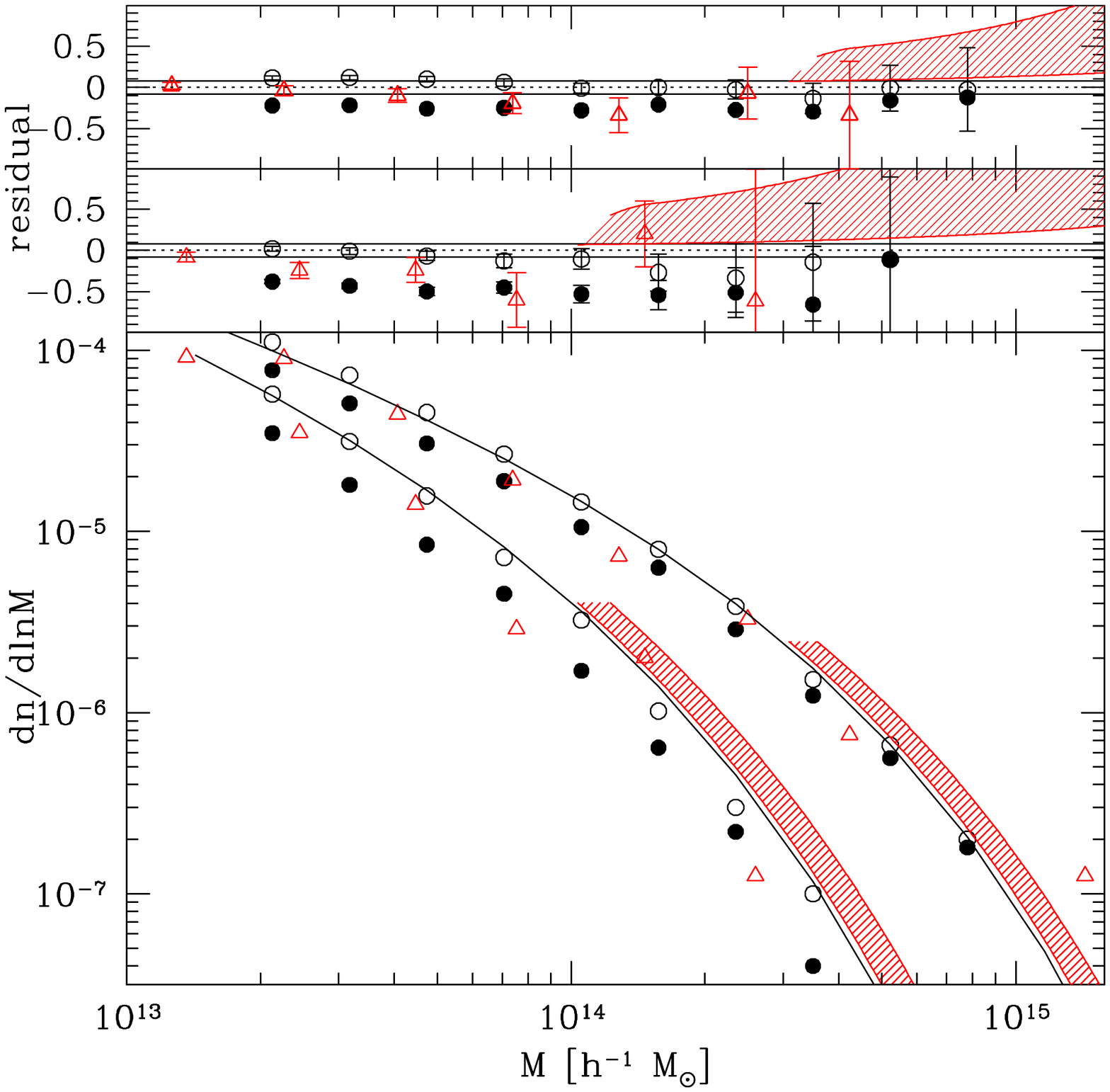}}
\caption{The lower panel shows the halo mass functions for the \PH (solid circles) and \GO (open circles) versions of the MGS and the ART--\GO model (open triangles) at redshifts $z = 0$ (upper) and $z=1$ (lower).  The  solid black lines are the TMF expectations at these redshifts.  Red bands shows the $90\%$ confidence regions anticipated by the shifts in halo mass for the ART--\CSF treatment.   The panels above show the fractional difference in number counts between the measured mass functions and the TMF, at $z=0$ (top) and $z=1$ (middle).  We plot $90\%$ jackknife errors for the ART--\GO model, and Poisson errors for the MGS.  Note that we have used the TMF for scaling the ART mass functions to match the MGS cosmology.}
\label{fig:nm}
\end{figure}

The shifts in mass seen with complex physical treatments will lead to changes in the mass function relative to the \GO models.  For both MGS simulations and the ART--\GO run, we compute binned space densities directly from the simulation counts.   Figure \ref{fig:nm} plots these mass functions at redshifts $z = 0$ and $z = 1$, and compares them to the TMF expectations for mean density contrasts equivalent to $\Delta_c = 500$, shown by the solid lines.  To account for differences in cosmology (primarily the difference in $\sigma_8$) we calculate the TMF for both cosmologies.  From a fixed number density, 
we find the mass shift between the two cosmologies, and apply it to the ART--\GO data for 
simple comparison with the MGS mass functions.

The redshift zero \GO mass functions match the TMF prediction quite well.  The top panels show the fractional difference in counts between the simulations and the TMF, with the $90\%$ statistical calibration uncertainty of the latter shown by the solid, horizontal lines.  We include $90\%$ uncertainties on the data points: jackknife uncertainties as a measure of cosmic variance for the ART--\GO sample, and Poisson uncertainties for the larger volume of the MGS simulation. The counts of both the ART and MGS models under \GO treatment lie within the TMF expectations at  $z=0$. 


At all masses, the \PH halo mass function is suppressed with respect to the \GO halo mass function, at a statistically significant level. At $M_{500} \sim 10^{14} \hinv \msol$, the number density of \PH halos is $20\%$ lower than the TMF prediction, a $4 \sigma$ shift relative to the $5\%$ TMF calibration error.   At the very high mass end, $\sim 10^{15}  M_\odot$, there is consistency with the TMF expectations.  

We do not have a complete mass function from the \CSF simulation.  However, we can anticipate the shift in halo number based on the mean shifts in halo mass presented above.  Since the ART--\GO models are consistent with the TMF,  we derive \CSF expectations by shifting the mass by fractional values given by the $90\%$ confidence range of the mean shifts shown in Fig~\ref{fig:dm}, meaning $\Delta M / M = 0.126 \pm 0.023$ at $z=1$ and $0.117 \pm 0.024$ at $z=0$.   We apply these shifts at mass scales probed by the \CSF halos, $M_{500} > 2 \times 10^{14} \hinv\msol$.   At $z=0$, the positive shift in halo mass  implies upward deviations in number density from the TMF expectation, at levels ranging from $10\%$ to $60\%$.  

\section{Discussion and Conclusion}\label{sec:conc}

Calibrations of the halo space density from ensembles of N-body and dissipationless gas dynamic simulations now have very small statistical uncertainties, $\sim 5\%$ in number \citep{tinker:08}.   
At the high-mass end, this level of precision in number is equivalent to a precision in halo mass at the $2\%$ level.   Since baryons represent $17\%$ of the matter density, complex gas dynamics associated with galaxy formation physics could plausibly lead to effects on halo masses of more than a few percent.  
In this letter, we demonstrate that shifts approaching $10\%$ in mass are possible, and that the sign of this effect is not yet understood.  

We use two extreme treatments of gas physics that are likely to bracket the range of behavior due to astrophysical processes in galaxy clusters.  A simple assumption of preheating reduces the local baryon fraction in halos, thereby suppressing their mass at levels ranging from $15\%$ at $10^{14} \hinv\msol$ to $5\%$ at $10^{15} \hinv\msol$.  A more complete physics treatment with cooling and star formation increases the local baryon fraction and deepens the halo potential, thus enhancing halo mass, by an average of $12\%$ at  $10^{14.5} \hinv\msol$.  The effects of cooling and star formation on halo mass are qualitatively consistent with the  systematic enhancement in small-scale power seen in previous simulations \citep{jing:06,rudd:08}.    In both of the complex physical treatments we consider, the shifts in mass lead to statistically significant offsets in cluster counts from the TMF expectations.  These shifts in mass depend on the choice of scale used in defining halos: in both treatments, the mass shifts are larger when identifying halos via higher density contrasts.

Although both the \PH and \CSF simulations provide fair matches to the mean observed $L-T$ relation, implying the structure of the hot gas phase is nearly correct, neither describes well the stellar content  of clusters.   The \PH simulation ignores galaxies while  the \CSF simulation converts nearly $\sim 50\%$ of baryons into a large stellar component.   Although it is tempting to dismiss the \PH model due to its lack of detailed physics, a growing body of observations, particularly the ubiquity of strong winds in moderate redshift DEEP2 galaxies \citep{weiner:08} and the remarkably simple evolution to $z=1.4$ of the color of red sequence galaxies seen in the Spiter/IRAC Shallow Survey \citep{eisenhardt:08}, provide supporting evidence for a scenario in which the fireworks associated with galaxy formation in clusters is both rapid and effective.  

Improvements in the physical and computational modeling of cooling and star formation are needed to match the full set of observational constraints on the baryonic mass components of cluster halos.  We have shown here that varying these treatments can affect total halo masses at levels up to ten percent.  
Improving the accuracy of the halo mass function calibration will therefore entail a suite of sophisticated  gas dynamic simulations, not more or larger N-body simulations.  

\section*{Acknowledgments}
We thank Elena Rasia, Daisuke Nagai, and Jeremy Tinker for their helpful comments.
This work was supported in part by NSF AST-0708150.  
DHR gratefully acknowledges the support of the Institute for Advanced
Study.  The ART simulations were performed on the Marenostrum
supercomputer at the Barcelona Supercomputing Center (BSC).  The MGS simulations 
were performed at Nottingham University, and we thank Frazer Pearce for providing 
the simulation data.



%
\newpage
%


\end{document}